\documentclass[preprint,showpacs,preprintnumbers,amsmath,amssymb]{article}
\usepackage{txfonts}
\usepackage{graphicx}
\usepackage{dcolumn}
\usepackage{bm}

\begin{document}

\title{Master Constraint Operators in Loop Quantum Gravity}

\author{Muxin Han$^{1,2}$\footnote{Email\ address:\
mhan1@lsu.edu} \ and
Yongge Ma$^{1}$\footnote{Email\ address:\ mayg@bnu.edu.cn}\\ \\
\small $1.$ Department of Physics, Beijing Normal University, \\
\small Beijing 100875, CHINA \\ \small $2.$ Horace Hearne Jr.
Institute for Theoretical Physics, \\ \small Louisiana State University, \\
\small Baton Rouge, LA 70803, USA}

\date{\today}

\maketitle

\begin{abstract}
We introduce a master constraint operator $\hat{\mathbf{M}}$ densely
defined in the diffeomorphism invariant Hilbert space in loop
quantum gravity, which corresponds classically to the master
constraint in the programme. It is shown that $\hat{\mathbf{M}}$ is
positive and symmetric, and hence has its Friedrichs self-adjoint
extension. The same conclusion is tenable for an alternative master
operator $\hat{\mathbf{M'}}$, whose quadratic form coincides with
the one proposed by Thiemann. So the master constraint programme for
loop quantum gravity can be carried out in principle by employing
either of the two operators.

\end{abstract}

{PACS number(s): 04.60.Pp, 04.60.Ds}

\section{Introduction}

It is well known that the quantization programme of loop quantum
gravity is based on the connection dynamics of general relativity
\cite{AL}\cite{rovelli}\cite{thiemann2}. The basic conjugate pairs
in the phase space are $su(2)$-valued connections $A^i_a$ and
densitized triads $\widetilde{P}^a_i$ on a 3-manifold $\Sigma$. In
the case where $\Sigma$ is a compact set without boundary, the
Hamiltonian is a linear combination of constraints as follows:
\begin{eqnarray}
H_{tot}=\mathcal{G}(\Lambda)+\mathcal{V}(\vec{N})+\mathcal{H}(N).
\end{eqnarray}
As an infinite dimensional Poisson algebra, the constraints algebra
is not a Lie algebra unfortunately, because the Poisson bracket
between the two scalar (Hamiltonian) constraints $\mathcal{H}(N)$
and $\mathcal{H}(M)$ has structure function depending on dynamical
variables \cite{AL}. This character causes much trouble in solving
the constraints quantum mechanically. On the other hand, the algebra
generated by the Gaussian constraints $\mathcal{G}(\Lambda)$ forms
not only a subalgebra but also a 2-side ideal in the full constraint
algebra. Thus one can first solve the Gaussian constraints
independently. But the subalgebra generated by the diffeomorphism
constraints $\mathcal{V}(\vec{N})$ can not form an ideal. Hence the
procedures of solving the diffeomorphism constraints and solving the
Hamiltonian constraints are entangled with each other. This leads to
certain ambiguity in the construction of a Hamiltonian constraint
operator \cite{thiemann1}\cite{GL}\cite{LM}. Thus, although the
kinematical Hilbert space $\mathcal{H}_{Kin}$ and the diffeomorphism
invariant Hilbert space $\mathcal{H}_{Diff}$ in loop quantum gravity
have been constructed rigorously \cite{ALM}, the quantum dynamics of
the theory is still an open issue. The regulated Hamiltonian
constraint operator $\hat{\mathcal{H}}^\epsilon(N)$ can be densely
defined in $\mathcal{H}_{Kin}$ and diffeomorphism covariant by
certain state-dependent triangulation $T(\epsilon)$, which may
naturally give a dual Hamiltonian constraint operator
$\hat{\mathcal{H}}'(N)$ acting on diffeomorphism invariant states
\cite{thiemann1}\cite{han2}. Moreover, one may even define a
symmetric version of regulated Hamiltonian constraint operator
\cite{thiemann5}. However, there are still several unsettled
problems concerning either form of the (dual) Hamiltonian constraint
operators, which are listed below.
\begin{itemize}

\item Although the action of the dual commutator of two
Hamiltonian constraint operators on
$\Psi_{Diff}\in\mathcal{H}_{Diff}$ reads
\begin{eqnarray}
([\hat{\mathcal{H}}(N),\hat{\mathcal{H}}(M)])'\Psi_{Diff}=0,
\end{eqnarray}
it is unclear whether the commutator between two Hamiltonian
constraint operators resembles the classical Poisson bracket
between two Hamiltonian constraints. Hence it is doubtful whether
the quantum Hamiltonian constraint produces the correct quantum
dynamics with correct classical limit \cite{GL}\cite{LM}, so it is
in danger of physical quantum anomaly.

\item Although the action of the dual commutator between the
Hamiltonian constraint operator and finite diffeomorphism
transformation operator $\hat{U}_\varphi$ on $\Psi_{Diff}$ gives
\cite{han2}
\begin{eqnarray}
([\hat{\mathcal{H}}(N),\hat{U}_\varphi])'\Psi_{Diff}
=\hat{\mathcal{H}}'(\varphi^*N-N)\Psi_{Diff},\label{anomalyfree}
\end{eqnarray}
which almost resembles the classical Poisson bracket between the
Hamiltonian constraint and diffeomorphism constraint, one can see
that the dual Hamiltonian constraint operator does not leave
$\mathcal{H}_{Diff}$ invariant. Thus the inner product structure
of $\mathcal{H}_{Diff}$ cannot be employed in the construction of
physical inner product.

\item Classically the collection of Hamiltonian constraints do not
form a Lie algebra. So one cannot employ group average strategy in
solving the Hamiltonian constraint quantum mechanically, since the
strategy depends on group structure crucially \cite{GM}.
\end{itemize}
However, if one could construct an alternative classical constraint
algebra, giving the same constraint phase space, which is a Lie
algebra (no structure function) and where the subalgebra of
diffeomorphism constraints forms an ideal, then the programme of
solving constraints would be much improved at a basic level. Such a
constraint Lie algebra was first introduced by Thiemann in
Ref.\cite{thiemann3}. The central idea is to introduce the master
constraint:
\begin{eqnarray}
\textbf{M}:=\frac{1}{2}\int_\Sigma
d^3x\frac{|\widetilde{C}(x)|^2}{\sqrt{|\det
q(x)|}},\label{mconstraint}
\end{eqnarray}
where the scalar constraint $\widetilde{C}(x)$ reads
\begin{eqnarray}
\widetilde{C}=\frac{\widetilde{P}^a_i\widetilde{P}^b_j}{\sqrt{|\det
q|}}[\epsilon^{ij}_{\ \ k}F^k_{ab}-2(1+\gamma^2)K^i_{[a}K^j_{b]}].
\end{eqnarray}
After solving the Gaussian constraint, one gets the master
constraint algebra as a Lie algebra:
\begin{eqnarray}
\{\mathcal{V}(\vec{N}),\ \mathcal{V}(\vec{N}')\}&=&\mathcal{V}([\vec{N},\vec{N}']),\nonumber\\
\{\mathcal{V}(\vec{N}),\ \textbf{M}\}&=&0,\nonumber\\
\{\textbf{M},\ \textbf{M}\}&=&0,\label{malgebra}
\end{eqnarray}
where the subalgebra of diffeomorphism constraints forms an ideal.
So it is possible to define a corresponding master constraint
operator on $\mathcal{H}_{Diff}$. A candidate self-adjoint master
constraint operator was first proposed in Ref.\cite{thiemann4}
from the positive quadratic form on $\mathcal{H}_{Diff}$
introduced in Ref.\cite{thiemann3}. In the following section, we
will construct two candidate self-adjoint master constraint
operators on $\mathcal{H}_{Diff}$ from a different perspective.
One of them coincides with the one proposed by Thiemann.

\section{Self-adjoint Master Constraint Operators}

We first introduce a master constraint operators densely defined in
${\cal H}_{Diff}$, then prove that it is symmetric and positive and
hence has its natural self-adjoint extension. Moreover, the Lie
algebra property of Eq.(\ref{malgebra}) is maintained in its quantum
version. The other master constraint operator can be constructed in
a similar way. The regularized version of the master constraint can
be expressed as
\begin{eqnarray}
\textbf{M}^{\epsilon}:=\frac{1}{2}\int_\Sigma d^3y \int_\Sigma
d^3x\chi_\epsilon(x-y)\frac{\widetilde{C}(y)}{\sqrt{V_{U_y^\epsilon}}}
\frac{\widetilde{C}(x)}{\sqrt{V_{U^\epsilon_{x}}}},
\end{eqnarray}
where $\chi_\epsilon(x-y)$ is any 1-parameter family of functions
such that $\lim_{\epsilon\rightarrow
0}\chi_\epsilon(x-y)/\epsilon^3=\delta(x-y)$ and
$\chi_\epsilon(0)=1$. Introducing a partition $\mathcal{P}$ of the
3-manifold $\Sigma$ into cells $C$, we have an operator
$\hat{H}^\epsilon_{C,\alpha}$ acting on any cylindrical function
$f_\alpha\in Cyl^3_\alpha(\overline{\mathcal{A}})$ in
$\mathcal{H}_{Kin}$ as
\begin{equation}
\hat{H}^\epsilon_{C,\alpha}\ f_\alpha=\sum_{v\in
V(\alpha)}\chi_C(v)\sum_{v(\Delta)=v}\frac{\hat{p}_{\Delta}}{\sqrt{\hat{E}(v)}}\hat{h}^{\epsilon,\Delta}_{v}
\frac{\hat{p}_{\Delta}}{\sqrt{\hat{E}(v)}}f_\alpha,\label{so1}
\end{equation}
via a state-dependent triangulation $T(\epsilon)$ on $\Sigma$, where
$\chi_C(v)$ is the characteristic function of the cell $C(v)$
containing a vertex $v$ of the graph $\alpha$, we use the
triangulation compatible with the symmetric Hamiltonian constraint
operator by asking the arcs $a_{ij}$ added by the Hamiltonian-like
operator $\hat{h}^{\epsilon,\Delta}_{v}$ to be smooth exceptional
edges defined in Ref.\cite{thiemann5}, and the tetrahedron projector
associated with segments $s_1$, $s_2$ and $s_3$ is
\begin{eqnarray}
\hat{p}_{\Delta}&:=&\hat{p}_{s_1}\hat{p}_{s_2}\hat{p}_{s_3}\nonumber\\
&=&\theta(\sqrt{\frac{1}{4}-\Delta_{s_1}}-\frac{1}{2})
\theta(\sqrt{\frac{1}{4}-\Delta_{s_2}}-\frac{1}{2})\theta(\sqrt{\frac{1}{4}-\Delta_{s_3}}-\frac{1}{2}),
\end{eqnarray}
here $\Delta_{s_i}$ is the Casimir operator associated with the
segment $s_i$ and $\theta$ is the distribution on $\mathbf{R}$ which
vanishes on $(-\infty,0]$ and equals $1$ on $(0,\infty)$, which
gives the vertex operator
$\hat{E}(v):=\sum_{v(\Delta)=v}\hat{p}_{\Delta}$. The expression of
$\hat{h}^{\epsilon,\Delta}_{v}$ reads
\begin{eqnarray}
\hat{h}^{\epsilon,\Delta}_{v}&=&\frac{8}{3i\hbar\kappa^2\gamma}\epsilon^{ijk}\mathrm{Tr}
\big(\{\hat{A}(\alpha_{ij}(\Delta))^{-1},\
\hat{A}(s_k(\Delta))^{-1}[\hat{A}(s_k(\Delta)),
\sqrt{\hat{V}_{U^\epsilon_{v}}}]\}\big)\nonumber\\
&&+2(1+\gamma^2)\frac{4\sqrt{2}}{3i\hbar^3\kappa^4\gamma^3}\epsilon^{ijk}\mathrm{Tr}\big(\hat{A}(s_i(\Delta))^{-1}
[\hat{A}(s_i(\Delta)),\hat{K}^\epsilon]\nonumber\\
&&\hat{A}(s_j(\Delta))^{-1}
[\hat{A}(s_j(\Delta)),\sqrt{\hat{V}_{U^\epsilon_{v}}}]\hat{A}(s_k(\Delta))^{-1}
[\hat{A}(s_k(\Delta)),\hat{K}^\epsilon]\big).\label{hath}
\end{eqnarray}
where $\{.,.\}$ denotes the anti-commutator,
$\{s_i(\Delta)\}_{i=1,2,3}$ is the segments associated to the
tetrahedron $\Delta$, and the arcs and segments constitute loops
$\alpha_{ij}(\Delta):=s_i(\Delta)\circ a_{ij}(\Delta)\circ
s_j(\Delta)$. Note that $\hat{h}^{\epsilon,\Delta}_{v}$ is similar
to that involved in the regulated symmetric Hamiltonian constraint
operator defined in Ref.\cite{thiemann5}, while the only difference
is that now the volume operator is replaced by its quare-root in
Eq.(\ref{hath}). Hence the action of $\hat{H}^\epsilon_{C,\alpha}$
on $f_\alpha$ adds smooth exceptional arcs $a_{ij}(\Delta)$ with
$\frac{1}{2}$-representation with respect to each $v(\Delta)$ of
$\alpha$. Thus, for each $\epsilon>0$, $\hat{H}^\epsilon_{C,\alpha}$
is a Yang-Mills gauge invariant and diffeomorphism covariant
operator defined on $Cyl^3_\alpha(\overline{\mathcal{A}})$. The
family of such operators with respect to different graphs is
cylindrically consistent up to diffeomorphisms and hence can give a
limit operator $\hat{H}_{C}$ densely defined on $\mathcal{H}_{Kin}$
by the uniform Rovelli-Smollin topology. Moreover, the regulated
operators are symmetric with respect to the inner product on
$\mathcal{H}_{Kin}$, i.e.,
\begin{eqnarray}
<g_{\gamma\ '},
\hat{H}^\epsilon_{C,\gamma}f_\gamma>_{Kin}=<\hat{H}^\epsilon_{C,\gamma}g_{\gamma\
'}, f_\gamma>_{Kin}=<\hat{H}^\epsilon_{C,\gamma\ '}g_{\gamma\ '},
f_\gamma>_{Kin},\label{symmetry}
\end{eqnarray}
which is non-vanishing provided $\gamma,\ \gamma\
'\in\Gamma_e(\gamma_0)$, where $\Gamma_e(\gamma_0)$ denotes the
collection of extended graphs obtained by adding only finite number
of smooth exceptional edges on an analytic skeleton $\gamma_0$.
Eq.(\ref{symmetry}) can be shown in analogy with the proof for
Theorem 3.1 in Ref.\cite{thiemann5} using the properties of smooth
exception edges. Then a master constraint operator,
$\hat{\mathbf{M}}$, acting on any $\Psi_{Diff}\in{\cal H}_{Diff}$
can be defined as:
\begin{equation}
(\hat{\textbf{M}}\Psi_{Diff})[f_\alpha]:=\lim_{\mathcal{P}\rightarrow
\Sigma;\epsilon,\epsilon'\rightarrow\mathrm{0}}\Psi_{Diff}[\sum_{C\in\mathcal{P}}
\frac{1}{2}\hat{H}^\epsilon_{C,\alpha}
\hat{H}^{\epsilon'}_{C,\alpha} f_\alpha].\label{master}
\end{equation}
Note that the actions of $\hat{H}^\epsilon_{C,\alpha}$ on any
$f_\alpha\in Cyl^3_\alpha(\overline{\mathcal{A}})$ only add finite
smooth exceptional arcs to the graph $\alpha$, and the newly added
vertices do not contribute in the successive action by the other
$\hat{H}^\epsilon_{C,\alpha}$. In addition,
$\hat{H}^\epsilon_{C,\alpha} \hat{H}^{\epsilon'}_{C,\alpha}
f_\alpha$ is a finite linear combination of spin-network functions
on an extended graph with the same analytic skeleton of $\alpha$,
hence the value of $(\hat{\textbf{M}}\Psi_{Diff})[f_\alpha]$ is
finite for any given $\Psi_{Diff}$. Thus
$\hat{\textbf{M}}\Psi_{Diff}$ lies in the algebraic dual
$\mathcal{D}^{*}$ of the space of cylindrical functions.
Furthermore, we can show that $\hat{\mathbf{M}}$ leaves the
diffeomorphism invariant distributions invariant. For any
diffeomorphism transformation $\varphi$ on $\Sigma$,
\begin{eqnarray}
(\hat{U}'_\varphi\hat{\textbf{M}}\Psi_{Diff})[f_\alpha]&=&\lim_{\mathcal{P}\rightarrow
\Sigma;\epsilon,\epsilon'\rightarrow\mathrm{0}}\Psi_{Diff}[\sum_{C\in\mathcal{P}}
\frac{1}{2}\hat{H}^\epsilon_{C,\varphi(\alpha)}
\hat{H}^{\epsilon'}_{C,\varphi(\alpha)}\hat{U}_\varphi
f_\alpha]\nonumber\\
&=&\lim_{\mathcal{P}\rightarrow
\Sigma;\epsilon,\epsilon'\rightarrow\mathrm{0}}\Psi_{Diff}[\hat{U}_\varphi\sum_{C\in\mathcal{P}}
\frac{1}{2}\hat{H}^{\varphi^{-1}(\epsilon)}_{\varphi^{-1}(C),\alpha}
\hat{H}^{\varphi^{-1}(\epsilon')}_{\varphi^{-1}(C),\alpha} f_\alpha]\nonumber\\
&=&\lim_{\mathcal{P}\rightarrow
\Sigma;\epsilon,\epsilon'\rightarrow\mathrm{0}}\Psi_{Diff}[\sum_{C\in\mathcal{P}}
\frac{1}{2}\hat{H}^\epsilon_{C,\alpha}
\hat{H}^{\epsilon'}_{C,\alpha} f_\alpha],
\end{eqnarray}
where in the last step, we used the fact that the diffeomorphism
transformation $\varphi$ leaves the partition invariant in the
limit $\mathcal{P}\rightarrow\Sigma$ and relabel $\varphi(C)$ to
be $C$. So we have the result
\begin{eqnarray}
(\hat{U}'_\varphi\hat{\textbf{M}}\Psi_{Diff})[f_\alpha]=(\hat{\textbf{M}}\Psi_{Diff})[f_\alpha].\label{diff}
\end{eqnarray}
Thus it is natural to define an inner product between
$\hat{\textbf{M}}\Psi_{Diff}$ and any diffeomorphism invariant
cylindrical function $\eta(f_\alpha)$ as
$<\hat{\textbf{M}}\Psi_{Diff}|\eta(f_\alpha)>_{Diff}:=(\hat{\textbf{M}}\Psi_{Diff})[f_\alpha]$.
Given any diffeomorphism invariant spin-network state $\Pi_{[s]}$
\cite{han2}\cite{thiemann2}\cite{AL}, the norm of the resulted state
$\hat{\textbf{M}}\Pi_{[s]}$ can be calculated as:
\begin{eqnarray}
||\hat{\textbf{M}}\Pi_{[s]}||_{Diff}&=&\sum_{[s']}|<\hat{\textbf{M}}\Pi_{[s]}|\Pi_{[s']}>_{Diff}|^2 \nonumber\\
&=&\sum_{[s']}|\lim_{\mathcal{P}\rightarrow
\Sigma;\epsilon,\epsilon'\rightarrow\mathrm{0}}\Pi_{[s]}[\sum_{C\in\mathcal{P}}
\frac{1}{2}\hat{H}^\epsilon_{C,\gamma(s')}
\hat{H}^{\epsilon'}_{C,\gamma(s')} \Pi_{s'\in[s']}]|^2\nonumber\\
&=&\sum_{[s']}|\lim_{\mathcal{P}\rightarrow
\Sigma;\epsilon,\epsilon'\rightarrow\mathrm{0}}\frac{1}{n_{\gamma(s)}}\sum_{\varphi\in
Diff/Diff_{\gamma(s)}}\sum_{\varphi'\in
GS_{\beta}}\nonumber\\
&&<\hat{U}_{\varphi}\hat{U}_{\varphi'}\Pi_{s\in[s]}
|\sum_{C\in\mathcal{P}}\frac{1}{2}\hat{H}^\epsilon_{C,\gamma(s')}
\hat{H}^{\epsilon'}_{C,\gamma(s')} \Pi_{s'\in[s']}>_{Kin}|^2\nonumber\\
&=&\sum_{[s']}|\lim_{\mathcal{P}\rightarrow
\Sigma;\epsilon,\epsilon'\rightarrow\mathrm{0}}\frac{1}{n_{\gamma(s)}}\sum_{\varphi\in
Diff/Diff_{\gamma(s)}}\sum_{\varphi'\in
GS_{\gamma(s)}}\nonumber\\
&&<\hat{U}_{\varphi}\hat{U}_{\varphi'}\sum_{C\in\mathcal{P}}\frac{1}{2}\hat{H}^{\epsilon'}_{C,\gamma(s)}
\hat{H}^{\epsilon}_{C,\gamma(s)} \Pi_{s\in[s]}
|\Pi_{s'\in[s']}>_{Kin}|^2\nonumber\\
&=&\sum_{[s']}|\lim_{\mathcal{P}\rightarrow
\Sigma;\epsilon,\epsilon'\rightarrow\mathrm{0}}\Pi_{[s']}[\sum_{C\in\mathcal{P}}
\frac{1}{2}\hat{H}^{\epsilon'}_{C,\gamma(s)}
\hat{H}^{\epsilon}_{C,\gamma(s)} \Pi_{s\in[s]}]|^2,\label{dense1}
\end{eqnarray}
where $n_\gamma$ is the number of the elements of the group,
$GS_\gamma$, of colored graph symmetries of $\gamma$,
$Diff_\gamma$ denotes the subgroup of $Diff$ which maps $\gamma$
to itself, $\gamma(s)$ is the graph associated with the
spin-network function $\Pi_s$, and we make use of the fact that
$\hat{\textbf{M}}$ commutes with diffeomorphism transformations.
The cylindrical function
$\sum_{C\in\mathcal{P}}\frac{1}{2}\hat{H}^{\epsilon'}_{C,\gamma(s)}
\hat{H}^{\epsilon}_{C,\gamma(s)} \Pi_{s\in[s]}$ is a finite linear
combination of spin-network functions on extended graph $\gamma\
'$ with the same analytic skeleton of $\gamma(s)$. Hence, fixing
$\Pi_{[s]}$, there are only finite number of terms which
contribute the sum in Eq.(\ref{dense1}). Thus the sum will
automatically converge. Note that one can give a more extensive
account of the terms contributing in Eq.(\ref{dense1}), in analogy
with the proof of the theorem 3.2 in Ref.\cite{thiemann4}.
Therefore, the master constraint operator $\hat{\mathbf{M}}$ is
densely defined on ${\cal H}_{Diff}$.

We now consider the property of $\hat{\mathbf{M}}$. Given two
diffeomorphism invariant spin-network functions $\Pi_{[s_1]}$ and
$\Pi_{[s_2]}$, the matrix elements of $\hat{\mathbf{M}}$ are
calculated as
\begin{eqnarray}
&&<\Pi_{[s_1]}|\hat{\textbf{M}}|\Pi_{[s_2]}>_{Diff}\nonumber\\
&=&\overline{(\hat{\textbf{M}}\Pi_{[s_2]})[\Pi_{s_1\in[s_1]}]}\nonumber\\
&=&\lim_{\mathcal{P}\rightarrow
\Sigma;\epsilon,\epsilon'\rightarrow\mathrm{0}}\sum_{C\in\mathcal{P}}\frac{1}{2}
\overline{\Pi_{[s_2]}[\hat{H}^\epsilon_{C,\gamma(s_1)}\hat{H}^{\epsilon'}_{C,
\gamma(s_1)}\Pi_{s_1\in[s_1]}]}\nonumber\\
&=&\lim_{\mathcal{P}\rightarrow
\Sigma;\epsilon,\epsilon'\rightarrow\mathrm{0}}\sum_{C\in\mathcal{P}}\frac{1}{2}
\frac{1}{n_{\gamma(s_2)}}\sum_{\varphi\in
Diff/Diff_{\gamma(s_2)}}\sum_{\varphi'\in
GS_{\gamma(s_2)}}\nonumber\\
&&\overline{<\hat{U}_{\varphi}\hat{U}_{\varphi'}\Pi_{s_2\in[s_2]}
|\hat{H}^\epsilon_{C,\gamma(s_1)}\hat{H}^{\epsilon'}_{C,\gamma(s_1)}\Pi_{s_1\in[s_1]}>_{Kin}}\nonumber\\
&=&\sum_{s}\lim_{\mathcal{P}\rightarrow
\Sigma;\epsilon,\epsilon'\rightarrow\mathrm{0}}\sum_{C\in\mathcal{P}}\frac{1}{2}
\frac{1}{n_{\gamma(s_2)}}\sum_{\varphi\in
Diff/Diff_{\gamma(s_2)}}\sum_{\varphi'\in
GS_{\gamma(s_2)}}\nonumber\\
&&\overline{<\hat{U}_{\varphi}\hat{U}_{\varphi'}\Pi_{s_2\in[s_2]}
|\hat{H}^\epsilon_{C,\gamma(s_1)}\Pi_s>_{Kin}<\Pi_s|\hat{H}^{\epsilon'}_{C,
\gamma(s_1)}\Pi_{s_1\in[s_1]}>_{Kin}}\nonumber\\
&=&\sum_{s}\lim_{\mathcal{P}\rightarrow
\Sigma;\epsilon,\epsilon'\rightarrow\mathrm{0}}\sum_{C\in\mathcal{P}}\frac{1}{2}
\frac{1}{n_{\gamma(s_2)}}\sum_{\varphi\in
Diff/Diff_{\gamma(s_2)}}\sum_{\varphi'\in
GS_{\gamma(s_2)}}\nonumber\\
&&\overline{<\hat{U}_{\varphi}\hat{U}_{\varphi'}\Pi_{s_2\in[s_2]}
|\hat{H}^\epsilon_{C,\gamma(s)}\Pi_s>_{Kin}<\Pi_s|\hat{H}^{\epsilon'}_{C,
\gamma(s_1)}\Pi_{s_1\in[s_1]}>_{Kin}}\nonumber\\
&=&\sum_{[s]}\sum_{v\in
V(\gamma(s\in[s]))}\frac{1}{2}\lim_{\epsilon,\epsilon'\rightarrow\mathrm{0}}
\overline{\Pi_{[s_2]}[\hat{H}^{\epsilon}_{v,\gamma(s)}
\Pi_{s\in[s]}]\sum_{s\in[s]}<\Pi_s|\hat{H}^{\epsilon'}_{v,\gamma(s_1)}\Pi_{s_1\in[s_1]}>_{Kin}},\nonumber\\
\label{element}
\end{eqnarray}
where we have used the resolution of identity trick in the fourth
step. Since only finite number of terms in the sum over
spin-networks $s$, cells $C\in\mathcal{P}$, and diffeomorphism
transformations $\varphi$ are non-zero respectively, we can
interchange the sums and the limit. In the fifth step, since only
the spin-network functions with
$\gamma(s),\gamma(s_1)\in\Gamma_e(\gamma_0)$ contribute the sum over
$s$, we can change $\hat{H}^\epsilon_{C,\gamma(s_1)}$ to
$\hat{H}^\epsilon_{C,\gamma(s)}$. In the sixth step, we take the
limit $C\rightarrow v$ and split the sum $\sum_s$ into
$\sum_{[s]}\sum_{s\in[s]}$, where $[s]$ denotes the diffeomorphism
equivalent class associated with $s$. Here we also use the fact
that, given $\gamma(s)$ and $\gamma(s')$ which are different up to a
diffeomorphism transformation, there is always a diffeomorphism
$\varphi$ transforming the graph associated with
$\hat{H}^{\epsilon}_{v,\gamma(s)} \Pi_s\ (v\in\gamma(s))$ to that of
$\hat{H}^{\epsilon}_{v',\gamma(s')} \Pi_{s'}\ (v'\in\gamma(s'))$
with $\varphi(v)=v'$, hence
$\Pi_{[s_2]}[\hat{H}^{\epsilon}_{v,\gamma(s)} \Pi_{s\in[s]}]$ is
constant for different $s\in[s]$.

Note that the term
$\sum_{s\in[s]}<\Pi_s|\hat{H}^{\epsilon'}_{v,\gamma(s_1)}\Pi_{s_1\in[s_1]}>_{Kin}$
in Eq.(\ref{element}) is free of the choices of the parameter
$\epsilon'$ up to diffeomorphisms. We thus use $[\epsilon']$ instead
of $\epsilon'$ to represent an arbitrary state-dependent
triangulation $T(\epsilon')$ in the diffeomorphism equivalent class.
Hence we get
\begin{eqnarray}
\sum_{s\in[s]}<\Pi_s|\hat{H}^{\epsilon'}_{v,\gamma(s_1)}\Pi_{s_1\in[s_1]}>_{Kin}
&=&\sum_{\varphi}<\hat{H}^{[\epsilon']}_{v,\varphi(\gamma(s))}U_\varphi\Pi_s|\Pi_{s_1\in[s_1]}>_{Kin}\nonumber\\
&=&\sum_{\varphi}<U_\varphi\hat{H}^{[\epsilon']}_{\varphi^{-1}(v),\gamma(s)}\Pi_s|\Pi_{s_1\in[s_1]}>_{Kin}\nonumber\\
&=&\overline{\Pi_{[s_1]}[\hat{H}^{\epsilon'}_{v\in
V(\gamma(s)),\gamma(s)}\Pi_s]},
\end{eqnarray}
where $\varphi$ are the diffeomorphism transformations spanning the
diffeomorphism equivalent class $[s]$. Note that the kinematical
inner product in above sum is non-vanishing if and only if
$\varphi(\gamma(s)),\gamma(s_1)\in\Gamma_e(\gamma_0)$ and $v\in
V(\varphi(\gamma(s)))$. Then the matrix elements (\ref{element}) are
resulted as:
\begin{eqnarray}
&&<\Pi_{[s_1]}|\hat{\textbf{M}}|\Pi_{[s_2]}>_{Diff}\nonumber\\
&=&\sum_{[s]}\sum_{v\in
V(\gamma(s\in[s]))}\frac{1}{2}\lim_{\epsilon,\epsilon'\rightarrow\mathrm{0}}
\overline{\Pi_{[s_2]}[\hat{H}^{\epsilon}_{v,\gamma(s)}
\Pi_{s\in[s]}]}\Pi_{[s_1]}[\hat{H}^{\epsilon'}_{v,\gamma(s)}\Pi_{s\in[s]}]\nonumber\\
&=&\sum_{[s]}\sum_{v\in
V(\gamma(s\in[s]))}\frac{1}{2}\overline{(\hat{H}'_v\Pi_{[s_2]})[
\Pi_{s\in[s]}]}(\hat{H}'_v\Pi_{[s_1]}) [
\Pi_{s\in[s]}].\label{master2}
\end{eqnarray}
From Eq.(\ref{master2}) and the fact that the master constraint
operator $\hat{\mathbf{M}}$ is densely defined on
$\mathcal{H}_{Diff}$, it is obvious that $\hat{\mathbf{M}}$ is a
positive and symmetric operator in ${\cal H}_{Diff}$. Therefore, the
quadratic form $Q_{\mathbf{M}}$ associated with $\hat{\mathbf{M}}$
is closable \cite{rs}. The closure of $Q_{\mathbf{M}}$ is the
quadratic form of a unique self-adjoint operator
$\hat{\overline{\mathbf{M}}}$, called the Friedrichs extension of
$\hat{\mathbf{M}}$. We relabel $\hat{\overline{\mathbf{M}}}$ to be
$\hat{\mathbf{M}}$ for simplicity. From the construction of
$\hat{\mathbf{M}}$, the qualitative description of the kernel of the
symmetric Hamiltonian constraint operator in Ref.\cite{thiemann5}
can be transcribed to describe the solutions to the equation:
$\hat{\mathbf{M}}\Psi_{Diff}=0$. In particular, the diffeomorphism
invariant cylindrical functions based on at most 2-valent graphs are
obviously normalizable solutions. In conclusion, there exists a
positive and self-adjoint operator $\hat{\mathbf{M}}$ on
$\mathcal{H}_{Diff}$ corresponding to the master constraint
(\ref{mconstraint}), and zero is in the point spectrum of
$\hat{\mathbf{M}}$.

Note that the quantum constraint algebra can be easily checked to be
anomaly free. Eq.(\ref{diff}) assures that the master constraint
operator commutes with finite diffeomorphism transformations, i.e.,
\begin{eqnarray}
[\hat{\mathbf{M}},\hat{U}'_\varphi]=0.
\end{eqnarray}
Also it is obvious that the master constraint operator commutes with
itself:
\begin{eqnarray}
[\hat{\mathbf{M}},\hat{\mathbf{M}}]=0.
\end{eqnarray}
So the quantum constraint algebra is consistent with the classical
constraint algebra (\ref{malgebra}) in this sense. As a result, the
difficulty of the original Hamiltonian constraint algebra can be
avoided by introducing the master constraint algebra, due to the Lie
algebra structure of the latter.

We notice that, similar to the non-symmetric Hamiltonian operator
\cite{thiemann1}, one can define a non-symmetric version of
Eq.(\ref{so1}) as
\begin{equation}
\hat{H'}^\epsilon_{C,\alpha}\ f_\alpha=\sum_{v\in
V(\alpha)}\chi_C(v)\sum_{v(\Delta)=v}\frac{\hat{p}_{\Delta}}{\sqrt{\hat{E}(v)}}\hat{h'}^{\epsilon,\Delta}_{v}
\frac{\hat{p}_{\Delta}}{\sqrt{\hat{E}(v)}}\
f_\alpha,
\end{equation}
where the operator $\hat{p}_{\Delta}/\sqrt{\hat{E}(v)}$ is suitably
arranged such that both $\hat{H'}^\epsilon_{C,\alpha}$ and its
adjoint are cylindrically consistent up to diffeomorphisms, and
\begin{eqnarray}
\hat{h'}^{\epsilon,\Delta}_{v}&=&\frac{16}{3i\hbar\kappa^2\gamma}\epsilon^{ijk}\mathrm{Tr}
\big(\hat{A}(\alpha_{ij}(\Delta))^{-1}\hat{A}(s_k(\Delta))^{-1}[\hat{A}(s_k(\Delta)),
\sqrt{\hat{V}_{U^\epsilon_{v}}}]\big)\nonumber\\
&&+2(1+\gamma^2)\frac{4\sqrt{2}}{3i\hbar^3\kappa^4\gamma^3}\epsilon^{ijk}\mathrm{Tr}\big(\hat{A}(s_i(\Delta))^{-1}
[\hat{A}(s_i(\Delta)),\hat{K}^\epsilon]\nonumber\\
&&\hat{A}(s_j(\Delta))^{-1}[\hat{A}(s_j(\Delta)),\hat{K}^\epsilon]\hat{A}(s_k(\Delta))^{-1}
[\hat{A}(s_k(\Delta)),\sqrt{\hat{V}_{U^\epsilon_{v}}}]\big)
\end{eqnarray}
via a state-dependent triangulation. The adjoint operator
$(\hat{H'}^{\epsilon}_{C,\alpha})^\dagger$ can be well defined in
${\cal H}_{kin}$ as
\begin{equation}
(\hat{H'}^\epsilon_{C,\alpha})^\dagger=\sum_{v\in
V(\alpha)}\chi_C(v)\sum_{v(\Delta)=v}\frac{\hat{p}_{\Delta}}{\sqrt{\hat{E}(v)}}(\hat{h'}^{\epsilon,\Delta}_{v})^\dagger
\frac{\hat{p}_{\Delta}}{\sqrt{\hat{E}(v)}},
\end{equation}
such that the limit operators $\hat{H'}_{C,\alpha}$ and
$(\hat{H'}_{C,\alpha})^\dagger$ in the uniform Rovelli-Smolin
topology satisfy
\begin{eqnarray}
<g_{\alpha'}, \hat{H'}_C f_{\alpha}>_{kin}&=&<g_{\alpha'},
\hat{H'}_{C,\alpha} f_{\alpha}>_{kin}=<(\hat{H'}_{C,\alpha})^\dagger
g_{\alpha'},f_{\alpha}>_{kin}\nonumber\\
&=&<(\hat{H'}_{C})^\dagger
g_{\alpha'},f_{\alpha}>_{kin}=<(\hat{H'}_{C})^\dagger_{\alpha'}g_{\alpha'},f_{\alpha}>_{kin},
\end{eqnarray}
where $\hat{H'}_C$ and $(\hat{H'}_{C})^\dagger$ are respectively the
inductive limits of $\hat{H'}_{C,\alpha}$ and
$(\hat{H'}_{C,\alpha})^\dagger$. Then an alternative master
constraint operator can be defined as \cite{han2}
\begin{equation}
(\hat{\mathbf{M'}}\Psi_{Diff})[f_\alpha]:=\lim_{\mathcal{P}\rightarrow
\Sigma;\epsilon,\epsilon'\rightarrow\mathrm{0}}\Psi_{Diff}[\sum_{C\in\mathcal{P}}\frac{1}{2}\hat{H'}^\epsilon_{C}
(\hat{H'}^{\epsilon'}_{C})^\dagger f_\alpha].
\end{equation}
In analogy with the previous discussion, we can show that
$\hat{\mathbf{M'}}$ is also qualified as a positive self-adjoint
operator on ${\cal H}_{Diff}$. Note that the construction of
$\hat{\mathbf{M'}}$ can be based only on the analytic category of
graphs. Moreover, the quadratic form of this operator coincides with
the quadratic form on (a dense form domain of) $\mathcal{H}_{Diff}$
defined by Thiemann in Ref.\cite{thiemann3}. Thus
$\hat{\mathbf{M'}}$ is equivalent to the master constraint operator
in Ref.\cite{thiemann4}.

\section{Discussions}

We have constructed two candidate self-adjoint master constraint
operators $\hat{\mathbf{M}}$ and $\hat{\mathbf{M'}}$ on ${\cal
H}_{Diff}$. As a candidate master operator $\hat{\mathbf{M}}$ is
different from that proposed in Ref.\cite{thiemann4}. In our
construction the structure of the kinematical Hilbert space
$\mathcal{H}_{Kin}$ is crucially employed, while in
Ref.\cite{thiemann4} the structure of the diffeomorphism invariant
Hilbert space ${\cal H}_{Diff}$ plays a key role. In
Ref.\cite{thiemann4}, the definition of the master operator
depends crucially on a quadratic form on ${\cal H}_{Diff}$, and a
new inner product is introduced on the algebra dual
$\mathcal{D}^{*}$ of the space of cylindrical functions in order
to have a well-defined quadratic form. Our construction shows
that, in a different perspective, master constraint operators can
be well defined on ${\cal H}_{Diff}$ without employing an inner
product on $\mathcal{D}^{*}$ and a quadratic form. Both approaches
can be used to construct master constraint operators for
background independent quantum matter fields coupled to gravity
\cite{HM}.

The aim of both Hamiltonian constraint programme and master
constraint programme is to seek for the physical Hilbert space
$\mathcal{H}_{phys}$. Since the master constraint operator
$\hat{\mathbf{M}}$ (or $\hat{\mathbf{M'}}$) is self-adjoint and a
separable $\mathcal{H}_{Diff}$ can be introduced by suitable
extension of diffeomorphism transformations \cite{FR}\cite{AL},
one can use the direct integral decomposition (DID) of
$\mathcal{H}_{Diff}$ associated with $\hat{\textbf{M}}$ to obtain
$\mathcal{H}_{phys}$ \cite{thiemann3}\cite{simon}. The physical
Hilbert space is just the (generalized) eigenspace of
$\hat{\textbf{M}}$ with the eigenvalue zero, i.e.,
$\mathcal{H}_{phys}=\mathcal{H}^\oplus_{\lambda=0}$ with the
induced physical inner product $<\ |\
>_{\mathcal{H}^\oplus_{\lambda=0}}$.
The issue of quantum anomaly is expected to be represented in terms
of the size of $\mathcal{H}_{phys}$ and the existence of sufficient
semi-classical states. The master constraint programme has been well
tested in various examples
\cite{thiemann9}\cite{DT2}\cite{DT3}\cite{DT4}\cite{DT5}. It is an
exciting result that the master constraint can be well defined as
self-adjoint operators in the framework of loop quantum gravity.
However, since the Hilbert spaces $\mathcal{H}_{Kin}$,
$\mathcal{H}_{Diff}$, and the master operators are constructed in
such ways that are drastically different from usual quantum field
theory, one has to check whether the constraint operators and the
corresponding algebra have correct classical limits with respect to
suitable semiclassical states. It is also possible to select a
preferred master operator from the alternative candidates by the
semiclassical analysis. However, to do the semiclassical analysis,
one still needs diffeomorphism invariant coherent states in
$\mathcal{H}_{Diff}$. The research in this aspect is now in progress
\cite{complexifier}\cite{ABC}.

Assume that the semiclassical analysis confirmed our master
constraint operator $\hat{\textbf{M}}$. Since $\hat{\textbf{M}}$ is
self-adjoint, it is a practical problem to find the DID of
$\mathcal{H}_{Diff}$ and the physical Hilbert space
$\mathcal{H}_{phys}$. However, the expression of master constraint
operator is so complicated that it is difficult to obtain the DID
representation of $\mathcal{H}_{Diff}$ directly. Fortunately, the
subalgebra generated by master constraints is an Abelian Lie algebra
in the master constraint algebra. So one can employ group averaging
strategy to solve the master constraint. Since $\hat{\textbf{M}}$ is
self-adjoint, by Stone's theorem there exists a strong continuous
one-parameter unitary group,
\begin{eqnarray}
\hat{U}(t):=\exp[it\hat{\textbf{M}}],
\end{eqnarray}
on $\mathcal{H}_{Diff}$. Then, given any diffeomorphism invariant
cylindrical functions $\Psi_{Diff}\in Cyl^\star_{Diff}$, one can
obtain algebraic distributions of $\mathcal{H}_{Diff}$ by a
rigging map $\eta_{phys}$ from $Cyl^\star_{Diff}$ to $Cyl_{phys}$,
which are invariant under the action of $\hat{U}(t)$ and
constitute a subset of the algebraic dual of $Cyl^\star_{Diff}$.
The rigging map is formally defined as
\begin{eqnarray}
\eta_{phys}(\Psi_{Diff})[\Phi_{Diff}]:=\int_\mathbf{R}\frac{dt}{2\pi}<\hat{U}(t)\Psi_{Diff}|\Phi_{Diff}>_{Diff}.
\end{eqnarray}
The physical inner product is then defined formally as
\begin{eqnarray}
&&<\eta_{phys}(\Psi_{Diff})|\eta_{phys}(\Phi_{Diff})>_{phys}:=\eta_{phys}(\Psi_{Diff})[\Phi_{Diff}]\nonumber\\
&&=\int_\mathbf{R}\frac{dt}{2\pi}<\hat{U}(t)\Psi_{Diff}|\Phi_{Diff}>_{Diff}.
\end{eqnarray}

\section*{Acknowledgments}

It is our pleasure to thank Professor Thomas Thiemann for his
advanced lectures at BNU and many enlightening discussions. The
authors would also like to thank the referee for helpful criticism
on the original manuscript. Muxin Han would like to acknowledge
all the members in the relativity group at BNU for all their kind
support. This work is supported in part by NSFC (10205002) and
YSRF for ROCS, SEM. Muxin Han would also like to acknowledge the
support from Undergraduate Research Foundation of BNU, the
fellowship and the assistantship of LSU, Hearne Foundation of LSU,
and funding from Advanced Research and Development Activity.


\begin{thebibliography}{99}

\bibitem{AL}A. Ashtekar and J. Lewandowski, Background independent quantum gravity: A status report,
Class. Quantum Grav. {\bf 21}, R53 (2004), (preprint:
gr-qc/0404018).

\bibitem{rovelli}C. Rovelli, Quantum Gravity, (Cambridge University Press,
2004).

\bibitem{thiemann2}T. Thiemann, Modern Canonical Quantum General
Relativity, (Cambrige Univeraity Press, in press).

\bibitem{thiemann1}T. Thiemann, Quantum spin dynamics (QSD), Class. Quantum Grav. {\bf
15}, 839 (1998).

\bibitem{GL}R. Gambini, J. Lewandowski, D. Marolf, and J. Pullin,
On the consistency of the constraint algebra in spin network
quantum gravity, Int. J. Mod. Phys. D {\bf 7}, 97 (1998).

\bibitem{LM}J. Lewandowski and D. Marolf, Loop constraints: a
habitate and their algebra, Int. J. Mod. Phys. D {\bf 7}, 299
(1998).

\bibitem{ALM}A. Ashtekar, J. Lewandowski, D. Marolf, J. Mour\~{a}o, and T.
Thiemann, Quantization of diffeomorphism invariant theories of
connections with local degrees of freedom, J. Math. Phys. {\bf
36}, 6456 (1995).

\bibitem{han2}M. Han, W. Huang, and Y. Ma, Fundamental structure of loop quantum gravity, preprint: gr-qc/0509064.

\bibitem{thiemann5}T. Thiemann, Quantum spin dynamics (QSD): II. The kernel of the Wheeler-DeWitt
constraint operator, Class. Quantum Grav. {\bf 15}, 875 (1998).

\bibitem{GM}N. Guilini and D. Marolf, On the generality of refined
algebraic quantization, Class. Quantum Grav. {\bf 16}, 2479
(1999).

\bibitem{thiemann3}T. Thiemann, The phoenix project: Master constraint programme
for loop quantum gravity, (preprint: gr-qc/0305080).

\bibitem{thiemann4} T. Thiemann, Quantum spin dynamics (QSD): VIII.
The master constraint, preprint: gr-qc/0510011.

\bibitem{rs} M. Reed and B. Simon, Methods of Modern Mathematical
Physics II: Fourer Analysis, Self-adjointness, (Academic Press,
1975), Page 177.

\bibitem{HM} M. Han and Y. Ma, Dynamics of scalar field in
polymer-like representation, Class. Quantum Grav. (in press),
preprint: gr-qc/0602101.

\bibitem{FR} W. Fairbain and C. Rovelli, Separable Hilbert space
in loop quantum gravity, J. Math. Phys. {\bf 45}, 2802 (2004).

\bibitem{simon}M. Reed and B. Simon, Methods of Modern Mathematical Phycics I: Functional
Analysis, (Academic Press, 1980).

\bibitem{thiemann9}B. Dittrich and T. Thiemann, Testing the master constraint programme for loop quantum gravity: I.
General framework, Class. Quantum Grav. {\bf 23}, 1025 (2006).

\bibitem{DT2}B. Dittrich and T. Thiemann, Testing the master constraint programme for loop quantum gravity: II.
Finite dimensional systems, Class. Quantum Grav. {\bf 23}, 1067
(2006).

\bibitem{DT3}B. Dittrich and T. Thiemann, Testing the master constraint programme for loop quantum gravity: III.
$SL(2,R)$ models, Class. Quantum Grav. {\bf 23}, 1089 (2006).

\bibitem{DT4}B. Dittrich and T. Thiemann, Testing the master constraint programme for loop quantum gravity: IV.
Free field theories, Class. Quantum Grav. {\bf 23}, 1121 (2006).

\bibitem{DT5}B. Dittrich and T. Thiemann, Testing the master constraint programme for loop quantum gravity: V.
Interecting field theories, Class. Quantum Grav. {\bf 23}, 1143
(2006).

\bibitem{complexifier}T. Thiemann, Complexifier coherent states for
quantum general relativity, (Preprint: gr-qc/0206037).

\bibitem{ABC}A. Ashtekar, L. Bombelli, and A. Corichi, Semiclassical states for constrained
systems, Phys. Rev. D {\bf 72}, 025008 (2005).


\end{thebibliography}
\end{document}